\documentclass[journal]{IEEEtran}
\usepackage{amsmath,amsfonts, amssymb, amsthm}
\usepackage{subcaption}  
\usepackage{algorithmic}
\usepackage{balance}
\usepackage[ruled,vlined]{algorithm2e}

\usepackage{amsthm}
\newtheorem{assumption}{Assumption}

\newtheorem{mydef3}{Lemma}

\usepackage{mathrsfs}
\usepackage{graphicx}
\usepackage{xcolor}
\usepackage{blindtext}

\usepackage{graphicx,epsf,epic,eepic,psfig,latexcad}

\usepackage{amssymb,amsmath,caption}
\usepackage{cite}
\usepackage{amsmath, amssymb, amsthm}
\usepackage{graphicx}

\usepackage{epstopdf}

\newcommand\numeq[1]%
{\stackrel{\scriptscriptstyle(\mkern-1.5mu#1\mkern-1.5mu)}{=}}
\makeatletter
\usepackage{textcomp}
\makeatother
\usepackage{xcolor}
\usepackage{url}

\begin{document}
	\title{Secret-Key-Based Physical Layer Security for Feedback-Aided Unsourced Random Access}

\author{Mohammad Javad Ahmadi, 
    Rafael F. Schaefer,~\IEEEmembership{Senior Member,~IEEE}, \\
    Farshad Rostami Ghadi,~\IEEEmembership{Member,~IEEE},
      and H. Vincent Poor,~\IEEEmembership{Life Fellow,~IEEE}
    \thanks{This work of M. J. Ahmadi and R. F. Schaefer have been supported in part by the German Federal Ministry of Research, Technology and Space (BMFTR) through the transfer hub \emph{6G-life} under Grant 16KIS2413K and in part by the German Research Foundation (DFG) as part of Germany's Excellence Strategy EXC 2050/2 -- Project ID 390696704 -- Cluster of Excellence \emph{``Centre for Tactile Internet with Human-in-the-Loop'' (CeTI)} and within the Priority Programme SPP 2378 ``Resilient Worlds'' -- Project ID 503657103. This work of F. Rostami Ghadi is supported by the European Union’s Horizon 2022 Research and Innovation Programme under Marie Skłodowska-Curie Grant No. 101107993. This work of H. V. Poor was supported in part by an Innovation Grant from Princeton NextG. An earlier version of this paper has been accepted for presentation in part at the \textit{2026 IEEE 102nd Vehicular Technology Conference (VTC2026-Spring)\cite{Ahmadi2025SURA}}.}
    \thanks{M. J. Ahmadi and R. F. Schaefer are with the Chair of Information Theory and Machine Learning and with the Cluster of Excellence \textit{``Centre for Tactile Internet with Human-in-the-Loop (CeTI),''} Technische Universit\"at Dresden, 01062 Dresden, Germany (e-mail: \{mohammad\_javad.ahmadi, rafael.schaefer\}@tu-dresden.de).}
    \thanks{F. Rostami Ghadi is with the Department of Signal Theory, Networking and Communications, Research Centre for Information and Communication Technologies (CITIC-UGR), University of Granada, 18071, Granada, Spain. (e-mail: f.rostami@ugr.es).}
    \thanks{H. V. Poor is with the Department of Electrical and Computer Engineering, Princeton University (e-mail: poor@princeton.edu).}
}

  \maketitle

\begin{abstract}
This work introduces security for unsourced random access (URA) via a physical-layer security approach. To achieve confidentiality, the proposed system opportunistically exploits intrinsic features of feedback-aided URA without altering its original structure or operational characteristics. As a result, the system preserves URA’s efficiency, including low delay and minimal signaling overhead, while ensuring secure communication. To secure transmission, each user generates a secret key from a feedback signal broadcast by the BS in a previous transmission round, which depends on the BS-user channel and can thus be treated as private. Each user then encrypts its data using the secret key before transmission. Along with the encrypted data, only the parity bits of the LDPC-encoded key are transmitted, enabling secret key recovery at the legitimate receiver via Slepian-Wolf decoding with side information. We propose a receiver algorithm to recover both the encrypted data and the encoded secret key at the legitimate receiver. We further present a theoretical analysis to derive analytical error probabilities for both the legitimate receiver and the passive eavesdropper, as well as to quantify the additional load imposed by the security measures on the URA system. It is shown, based on both theoretical analysis and simulation results, that meaningful secrecy is achieved with only negligible extra overhead compared to the standard URA system.
\end{abstract}
	\begin{figure}
		\centering
		\includegraphics[width=.9\linewidth, trim=110 1 280
		70, clip]{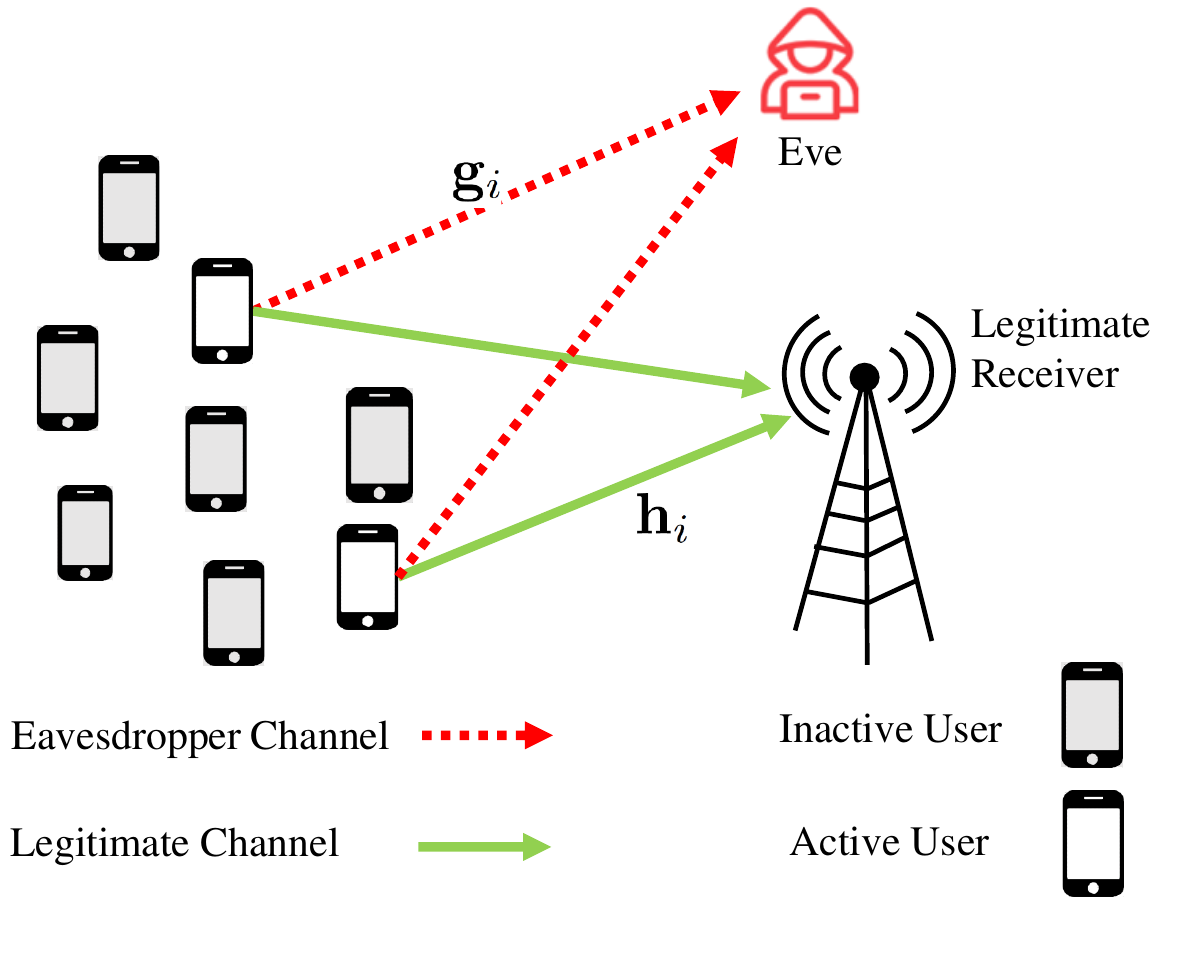}
\caption{{\small Secure unsourced random access model in which only a small subset of users is active at a given time. Signals from each active user are received by the legitimate base station and by a passive eavesdropper over distinct channels.}}
		\label{Fig_configuratin}
	\end{figure}

	\vspace{-.8mm}
    
	\vspace{-.8mm}
	\section{Introduction}
In coordinated or grant-based multiple access systems, a user must first complete an access procedure involving several signaling steps for identification and registration. After that, the base station (BS) performs additional signaling to allocate dedicated transmission resources to the user~\cite{Rappaport}. However, performing several rounds of interaction with each user substantially increases the system complexity, especially in scenarios with a large number of active devices or rapidly changing traffic conditions~\cite{Ozates2024Survey}. To reduce signaling overhead in large-scale systems, \textit{unsourced random access} (URA) was introduced in~\cite{polyanskiy2017perspective}. In this paradigm, users transmit immediately upon having data to send, without prior coordination with the BS for identification or resource allocation. Consequently, user identities are not revealed, motivating the term “unsourced,” while the lack of scheduling-based resource allocation motivates the term “random access”~\cite{Ahmadi2024RISUMA}. Although unsourced and random access features offer low latency and reduced signaling overhead by eliminating scheduling and user identification, the system faces significant challenges such as collisions and interference, which must be resolved at the physical layer~\cite{Ozates2024Survey}. These challenges make signal recovery at the receiver more difficult and require high computational complexity, which grows at least linearly with the number of active users~\cite{Ahmadi2023Unsourced}. Therefore, the main task in URA is to separate signals from unidentified users and recover them with minimal computational complexity. Despite numerous works investigating algorithmic and theoretical aspects of URA systems~\cite{Ahmadi2023Unsourced,andreev2020polar,nassaji2022spread,Zhang2025Sparse,ahmadi2021random,Zhang2024Dictionary,polyanskiy2017perspective,Ahmadi2024RISUMA,Zhang2024Blind,Ahmadi2024TWCIntegrated,Ahmadi2024Integrated}, no work has addressed secure transmission for the users, leaving URA systems vulnerable to eavesdropping and data leakage. 

Physical layer security (PLS) offers a lightweight alternative to classical cryptographic methods by leveraging the inherent properties of the physical communication medium to ensure confidentiality~\cite{Rafael2017}. Unlike computational cryptography, which relies on the assumed hardness of mathematical problems and requires significant computational resources for encryption, key distribution, and authentication, PLS can secure transmissions without imposing heavy processing overhead on devices. This characteristic is particularly advantageous for systems composed of low-cost or resource-constrained devices, such as  those envisioned in large-scale 6G deployments like URA. In such systems, not only are signals transmitted openly and can be received by any device within range, but the large number of connected devices also amplifies the computational burden of traditional cryptographic measures~\cite{Sun2022PLS}. \textcolor{black}{Among the various techniques in PLS, secret key generation from wireless channels has gained significant attention, as it enables legitimate users to establish shared secret keys by exploiting the inherent randomness and reciprocity of the wireless medium~\cite{Mathur2008,Ren2011,Baksi2017,Sun2010,Gunlu2023,Islam2015,Schaefer2018,Gunlu2018}.}

\indent In the PLS literature, users and the BS often exchange pilot or training signals to estimate the channel and generate secret keys from the observed channel conditions~\cite{Rafael2017}. Such schemes can introduce significant signaling overhead, especially in large-scale networks with many users. In contrast, in this paper, we propose a secure system, termed secure URA (SURA), in which security measures are incorporated into a feedback-aided URA system without compromising its unsourced and random access characteristics. In SURA, the system opportunistically leverages the existing URA configuration in two ways: (1) new users exploit the downlink signal transmitted by the BS as a private observation, which is broadcast for general purposes such as feedback to previous users or sensing nearby targets~\cite{Ahmadi2025Feedback,Ebert2022}; and (2) the system reuses the channel estimation already performed in conventional URA schemes for channel estimation, user separation, and decoding. Importantly, these two components are already present in the original feedback-aided URA and are used in SURA solely to enable secure transmission without introducing additional signaling or processing overhead. This opportunistic use of existing features in URA, enables SURA to integrate security into URA while maintaining rapid access, low signaling overhead, and minimal latency.

\vspace{1mm}
Our contributions are as follows:\\
\noindent \textbf{Secure transceiver design}: We develop a complete design for SURA, detailing both transmitter and receiver algorithms:
\begin{itemize}
        \item Transmitter design: Each new user first receives the downlink signal from the BS, which is used to generate its secret key. The user then encrypts its data using the generated secret key. Next, the user encodes and modulates both the encrypted data and the secret key. It then appends the modulation of the encrypted data to the parity part of the modulated secret key codeword before transmission.

\item Receiver design: At the receiver, the BS recovers the secret key and then uses it to decrypt the user data. To recover the secret key, the BS employs a Slepian-Wolf decoding method, which combines the parity part of the encoded key received over the uplink with side information obtained from estimating each user's observation of the downlink signal. {\color{black} The security of the generated secret key relies on the statistical independence between the user's observation and that of a passive eavesdropper due to independent fading channels~\cite{Baksi2017,Mathur2008,Ren2011}. Moreover, the LDPC-based Slepian--Wolf reconciliation reveals only limited information about the key to the eavesdropper, which is insufficient for reliable recovery in the absence of the corresponding side information available at the BS~\cite{Baksi2017,Sun2010,Gunlu2023,Islam2015}.}
\end{itemize}

\noindent \textbf{Theoretical analysis}: We develop a comprehensive information-theoretic analysis of the SURA system:
\begin{itemize}
    \item For each user, we derive the information density at the legitimate receiver under favorable assumptions and apply a Gaussian approximation to obtain a lower bound on the user’s block error probability. Averaging over all active users yields the system’s overall per-user probability of error (PUPE).

    \item Using the same methodology, we derive a lower bound on the PUPE at a passive eavesdropper, allowing us to quantify the secrecy level of each transmission.

    \item Finally, we analyze the additional resource and processing overhead introduced by the security measures, providing benchmarks for SURA’s performance, secrecy, and overhead relative to a standard feedback-aided URA system.
\end{itemize}
\indent The rest of the paper is organized as follows. Sec.~\ref{Sec_SysModel} presents the system model. Sec.~\ref{Sec_poposdScheme} introduces the proposed SURA scheme, detailing the transmitter and receiver design. Sec.~\ref{SecTheoretical} provides a theoretical analysis of the system’s performance, secrecy, and overhead. Sec.~\ref{Numerical} presents numerical results illustrating the effectiveness of the proposed scheme. Finally, Sec.~\ref{concluson} concludes the paper.
	\section{System Model}
    \label{Sec_SysModel}

We consider a feedback-aided communication system with a BS equipped with $M$ receive antennas, and a massive number of connected users among which $K_a$ users with indices $i=1,2,\ldots,K_a$ aim to securely share their bit sequences $\mathbf{w}_i\in\{0,1\}^{B}$ with the legitimate BS. A passive Eve equipped with $E$ receive antennas attempts to illegally intercept the signals transmitted by the users. When the $i$th user is ready to transmit its information, it waits for a feedback signal from the BS, which is a non-user-specific broadcast signal frequently sent for communication and sensing purposes~\cite{Ahmadi2025Feedback,Ebert2022}. Considering channel reciprocity, the feedback signal received by the $i$th user is given by~\cite{Ahmadi2025Feedback}
	\begin{align}
		\mathbf{y}_{i} = \mathbf{h}_i^T \mathbf{V}+\mathbf{o}_{i},\label{Eq_downlink}
	\end{align}
where $\mathbf{h}_i \in \mathbb{C}^{M\times 1}$ denotes the channel vector from user $i$ to the BS, $\mathbf{V} \in \mathbb{C}^{M \times L}$ is the downlink signal transmitted from the $M$ BS antennas over $L$ channel uses, each column of which has a power norm of $P_f M$. The term $\mathbf{o}_i \in \mathbb{C}^{1 \times L}$ represents the additive white Gaussian noise (AWGN) vector, where each entry is drawn from $\mathcal{CN}(0,\sigma_u^{2})$. If the receiver has access to an estimate of the channel coefficient vector \(\hat{\mathbf{h}}_i\), since it perfectly knows the downlink signal \(\mathbf{V}\) that it has sent, it can obtain an estimate of \(\mathbf{y}_i\) as
\begin{align}
	\hat{\mathbf{y}}_i = \hat{\mathbf{h}}_i^T \mathbf{V}.\label{Eq_est_yi}
\end{align}
For the passive Eve, assuming it is located sufficiently far from the \(i\)th user, the vector \(\mathbf{y}_i\) is effectively independent of Eve’s observation. Hence, \(\mathbf{y}_i\) can serve as a private observation vector that can be exploited as a source of security between the \(i\)th user and the legitimate BS.

Assuming synchronous transmission, the received signals at the legitimate BS and at the Eve are expressed as
	\begin{subequations}
	\begin{align}
		\mathbf{Y}_{\mathrm{BS}} = \sum_{i=1}^{K_a}\mathbf{h}_i\mathbf{x}_i+\mathbf{Z},\label{Eq_uplink_bs}\\
		\mathbf{Y}_{\mathrm{Eve}} = \sum_{i=1}^{K_a}\mathbf{g}_i\mathbf{x}_i+\mathbf{N},\label{Eq_uplink_eav}
	\end{align}
	\label{eq_receivedSignal}
\end{subequations}
\noindent where $\mathbf{h}_i \in \mathbb{C}^{M \times 1}$ and $\mathbf{g}_i \in \mathbb{C}^{E \times 1}$ represent the channel vectors from the $i$th user to the BS and Eve, respectively, the vector $\mathbf{x}_i \in \mathbb{C}^{1\times n}$ is the transmit signal of the $i$th user, generated from its message sequence $\mathbf{w}_i$ and the received downlink signal $\mathbf{y}_i$. The matrices $\mathbf{Z}$ and $\mathbf{N}$ represent AWGN components, with independent entries distributed as $\mathcal{CN}(0, \sigma_c^2)$ and $\mathcal{CN}(0, \sigma_e^2)$, respectively.

We note that most physical layer security approaches rely on identified users with known channel state information, assumptions that do not hold in URA systems. In contrast, the proposed scheme enables secure data transmission in a fully unsourced and random access manner, without requiring any explicit signal exchange between the legitimate BS and the users for security purposes. As a result, the desirable features of URA, such as low delay and low signaling overhead, are preserved while providing secure communication.

\section{Proposed SURA Scheme}
\label{Sec_poposdScheme}
The overall procedure of the proposed secure scheme is summarized as follows. Each user first waits to receive the downlink signal from the BS, $\mathbf{y}_i$, and uses it to generate its secret key. The user then encrypts its message sequence using this key and transmits the encrypted data along with the encoded secret key. At the receiver, the BS recovers each user’s secret key and encrypted data, and subsequently decrypts the data to recover the original messages. With this overview, we proceed to describe the transmitter and receiver designs in detail.
\subsection{Transmitter Signal Design}
The transmit signal of each user, \(\mathbf{x}_i\), consists of three segments: the pilot segment, the polar segment, and the key segment. The key segment securely conveys the secret key, while the pilot and polar segments jointly enable the BS to perform pilot detection, channel estimation, and decoding of the encrypted bits. The details of each segment are provided below.

\subsubsection{Key Segment}
\label{Sec_transmit_Key}
The $i$th user projects its private observation $\mathbf{y}_i$ onto a vector of length $S$ as
\begin{align}
    \mathbf{u}_i = \left[\Re\{\mathbf{y}_i \mathbf{C}_1\},\Im\{\mathbf{y}_i \mathbf{C}_1\}\right] \in \mathbb{R}^{1 \times S},\label{eq_ui}
\end{align}
where \(\mathbf{C}_1 \in \mathbb{C}^{L \times 0.5S}\) is a projection matrix with orthonormal columns, i.e., \(\mathbf{C}_1^H \mathbf{C}_1 = \mathbf{I}_{0.5S}\). We note that for this condition to be satisfied, it is required that \(L \ge 0.5S\). To generate the secret key, each component of \(\mathbf{u}_i\) is quantized as
\begin{align}
    \mathbf{s}_i = F(\mathbf{u}_i),\label{eq_secretKey}
\end{align}
where \(F(\cdot)\) maps each entry of its input vector to \(0\) if it is negative and to \(1\) otherwise, producing a vector of the same length as its input. The \(i\)th user then encodes its secret key \(\mathbf{s}_i\) using a systematic \((n_s, S)\) low-density parity-check (LDPC) code. Owing to the systematic structure of the code, the resulting length-\(n_s\) codeword can be partitioned into two disjoint parts of lengths \(S\) and \(n_s - S\). Specifically, the first part, the systematic part, corresponds directly to the secret key \(\mathbf{s}_i\), while the second part, \(\tilde{\mathbf{s}}_i \in \{0,1\}^{n_s - S}\), contains the parity bits generated via the LDPC parity-check matrix.

 In the considered protocol, only the parity subvector $\tilde{\mathbf{s}}_i$ is transmitted, while the systematic part $\mathbf{s}_i$ is not directly included in the transmit signal. This choice is motivated by the presence of side information at the legitimate receiver (an \textit{a priori} estimate of $\mathbf{s}_i$ obtained from \(\hat{\mathbf{y}}_i\) in \eqref{Eq_est_yi}), which enables the receiver to perform {syndrome-based reconciliation} (Slepian–Wolf decoding) to recover $\mathbf{s}_i$. Transmitting only the parity (syndrome) reduces the information leaked to the Eve, while still allowing reliable recovery at the intended receiver. To generate the key segment of the transmit signal, the parity bits $\tilde{\mathbf{s}}_i$ are modulated using binary phase-shift keying (BPSK), mapping $0\mapsto+\sqrt{P_k}$ and $1\mapsto-\sqrt{P_k}$, yielding the transmitted signal in the key segment as
\begin{align}
    \mathbf{x}_{k,i} \in \left\{\pm \sqrt{P_k}\right\}^{n_s-S},\label{eq_keysegmentss}
\end{align}
where \(P_k\) is the per-channel-use power of the key segment.  

\subsubsection{Pilot and Polar Segments}
To ensure data confidentiality, the $i$th user encrypts its bit sequence $\mathbf{w}_i\in \{0,1\}^B$ using its secret key $\mathbf{s}_i \in \{0,1\}^{S}$ obtained in~\eqref{eq_secretKey}. Specifically, a keystream $\mathbf{k}_i$ of length $B$ bits is generated from $\mathbf{s}_i$ as
\begin{align}
	\mathbf{k}_i = \mathbf{s}_i\mathbf{T} \bmod 2,\label{eq_extendingTheKey}
\end{align}
where $\mathbf{T} \in \{0,1\}^{S \times B}$ is a publicly known binary matrix whose entries are independently drawn according to a Bernoulli$(1/2)$ distribution. This random construction ensures that, with high probability, the mapping $\mathbf{s}_i \mapsto \mathbf{k}_i$ is injective and induces a uniform ensemble over a subspace of size $2^S$. Consequently, the generated keystreams are statistically symmetric and uniformly distributed over their image, which plays a key role in establishing the intrinsic ambiguity at the eavesdropper. The ciphertext is then obtained by a bitwise XOR operation between the bit sequence and the keystream, i.e.,
\begin{equation}
	\mathbf{c}_i = \mathbf{w}_i \oplus \mathbf{k}_i.\label{eq_encrypted}
\end{equation}
Then the ciphertext $\mathbf{c}_i$ is divided into pilot and polar sub-messages as
\begin{align}
    \mathbf{c}_i=[\mathbf{c}_{p_i},\mathbf{c}_{d_i}],\label{eq_appendedbits}
\end{align}
where $\mathbf{c}_{p_i}\in \{0,1\}^{B_p}$ and $\mathbf{c}_{d_i}\in \{0,1\}^{B-B_p}$. The pilot sub-message $\mathbf{c}_{p_i}$ is mapped to the pilot codebook $\mathbf{P}\in \mathbb{C}^{2^{B_p}\times n_p}$ to generate its pilot segment, where the elements of $\mathbf{P}$ are randomly drawn from $\mathcal{CN}(0,1)$ and each row of it is normalized to satisfy $\|\mathbf{p}_i\|^2=n_pP_p$, where $P_p$ denotes the per-channel-use transmit power of the pilot segment, and $\mathbf{p}_i$ is the $i$th row of $\mathbf{P}$. Assuming without loss of generality that the $i$th user selects the $i$th row of the codebook, its pilot segment is written as  
\begin{align}
    \mathbf{x}_{p,i}=\mathbf{p}_i\in \mathbb{C}^{1\times n_p}.\label{eq_pilot}
\end{align}
Then, the polar sub-message $\mathbf{c}_{d_i}$ is appended by $B_r$ cyclic redundancy check (CRC) bits, then passed to an $\left(n_d, \  B - B_p +B_r\right)$ polar code, and the result is modulated using BPSK to construct the polar segment of the transmit signal
\begin{align}
   \mathbf{x}_{d,i}\in \left\{\pm \sqrt{P_d}\right\}^{n_d},\label{eq_data}
\end{align}
where $P_d$ represents the per-channel-use power of the polar segment. Finally, the full transmit signal of the $i$th user is generated by appending key, pilot, and polar segments in \eqref{eq_keysegmentss}, \eqref{eq_pilot}, and \eqref{eq_data} as
\begin{align}
    \mathbf{x}_i=[\mathbf{x}_{k,i},\mathbf{x}_{p,i},\mathbf{x}_{d,i}].\label{eq_transmitStrc}
\end{align}
\begin{figure}
	\centering
	\includegraphics[width=1\linewidth, trim=18 5 38 21, clip]{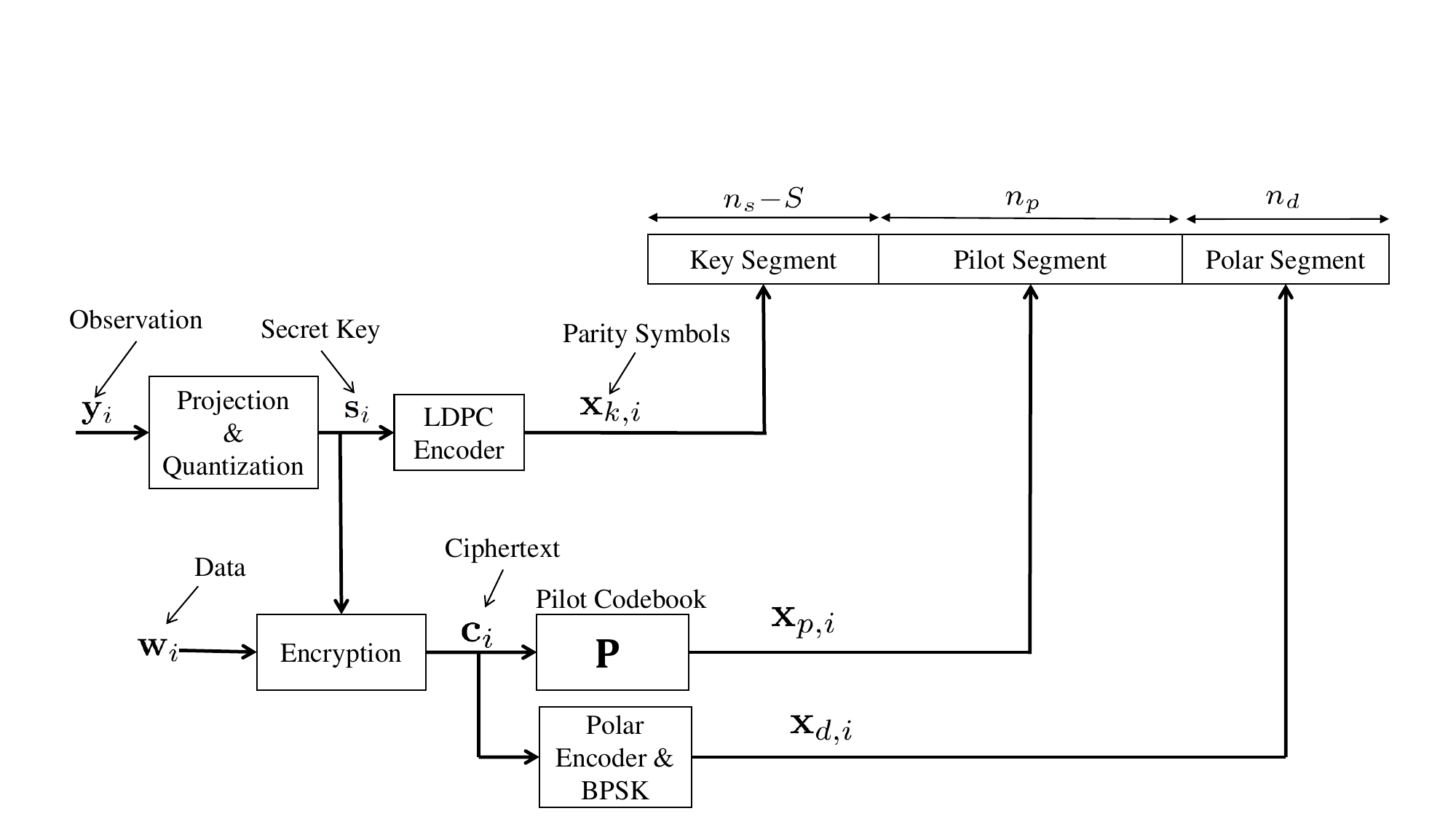}
\caption{{\small Block diagram of the transmit signal design, showing the generation of different segments: pilot, polar, and key segments.}}
	\label{transmitter}
\end{figure}

The procedure for generating the transmit signal of the $i$th user is summarized in Algorithm~\ref{algorithm1} and illustrated in Fig.~\ref{transmitter}.
\begin{algorithm}
  \caption{Transmitter} 
  \label{algorithm1}
 \textbf{Key Segment}
\begin{itemize}
     \item Generate secret key $\mathbf{s}_i$ in \eqref{eq_secretKey}.
\item Encode the secret key using LDPC, apply BPSK modulation, and extract the parity part to obtain the key segment $\mathbf{x}_{k,i}$ in \eqref{eq_keysegmentss}. \end{itemize}
\textbf{Pilot \& Polar Segments}
\begin{itemize}
  \item  Encrypt the users' data as in \eqref{eq_encrypted} to obtain $\mathbf{c}_i$.
    \item Map the first $B_p$ bits of $\mathbf{c}_i$ to the rows of $\mathbf{P}$ to obtain pilot segment $\mathbf{x}_{p,i}$ in \eqref{eq_pilot}.
\item The remaining $B - B_p$ bits of $\mathbf{c}_i$ are encoded with a polar code and then modulated to form the polar segment $\mathbf{x}_{d,i}$ in~\eqref{eq_data}.
\end{itemize}
Finally, the whole transmit signal of the $i$th user is obtained by appending $\mathbf{x}_{p,i},\mathbf{x}_{d,i},\mathbf{x}_{k,i}$ as in \eqref{eq_transmitStrc}. 
\end{algorithm}
\subsection{Legitimate Receiver Design}
As shown in \eqref{Eq_uplink_bs}, the signal received by the legitimate BS is the summation of the signals from \(K_a\) users whose identities are completely unknown to any receiver. To recover each user's signal, which is perturbed by interference from other users, the process is divided into two stages. In the first stage, an iterative algorithm is jointly applied to the pilot and polar segments of the received signal to recover each user’s ciphertext by detecting its pilot, estimating its corresponding channel coefficient vector, and decoding its polar segment. In the second stage, the received signal in the key segment is employed to recover each user's secret key. Finally, the recovered secret key is used to decrypt the ciphertext, thereby recovering the original message.

Focusing on the transmit signal structure in \eqref{eq_transmitStrc}, the received signal model in \eqref{Eq_uplink_bs} can be written as
\begin{subequations}
\begin{align}
    \mathbf{Y}_{\mathrm{BS}} &= \left[\mathbf{Y}_p,\mathbf{Y}_d,\mathbf{Y}_k\right]
    ,\label{Eq_uplink_bsMatrix}\\
    \mathbf{Y}_j&=\mathbf{H}\mathbf{X}_j+\mathbf{Z}_j, \quad j\in\{p,d,k\},\label{Eq_segment_Y}
\end{align}
\end{subequations}
where columns of $\mathbf{H}$ are $\mathbf{h}_i$'s for different users, rows of $\mathbf{X}_p$, $\mathbf{X}_d$, and $\mathbf{X}_k$ are $\mathbf{x}_{p,i}$, $\mathbf{x}_{d,i}$, and $\mathbf{x}_{k,i}$, shown in \eqref{eq_pilot}, \eqref{eq_data}, and \eqref{eq_keysegmentss}, respectively, and $\mathbf{Z}_j$ is the submatrix of $\mathbf{Z}$ corresponding to different segments.

\subsubsection{Iterative Algorithm}
\label{Sec_OMP}
The following three steps are performed to recover the encrypted messages of each user.

\noindent \textbf{Step~1 (pilot detection and channel estimation):}
 From \eqref{Eq_segment_Y}, the received signal matrix corresponding to the pilot segment is written as $\mathbf{Y}_p = \mathbf{H}\mathbf{X}_p+\mathbf{Z}_p$. Feeding the pilot codebook $\mathbf{P}$ and the received signal  $\mathbf{Y}_p$ into the orthogonal matching pursuit (OMP), a set of pilot sub-messages are detected and their channel coefficient vectors are estimated, i.e., $\hat{\mathbf{c}}_{p_i}$ and $\hat{\mathbf{h}}_i$ with $i\in \mathcal{D}$, where $\mathcal{D}$ is the set of detected pilot indices.

\noindent \textbf{Step~2 (polar decoding):}
In this part, we use the data segment of the received signal, $\mathbf{Y}_d$ in~\eqref{Eq_segment_Y}, to decode the polar bits $\mathbf{c}_{d_i}$ of each user. To this end, we apply least squares (LS) to obtain a soft estimation of $\mathbf{X}_d$ as
\begin{align}
	\hat{\mathbf{X}}_{d} =  \Re\left\{ (\hat{\mathbf{H}}^H\hat{\mathbf{H}})^{-1}\hat{\mathbf{H}}^H\mathbf{Y}_d\right\},\label{eq_LS}
\end{align}
where $\hat{\mathbf{h}}_i$ constitutes the $i$th column of $\hat{\mathbf{H}}$. Assuming perfect channel estimation, and focusing on the structure of $\mathbf{Y}_d$ in \eqref{Eq_segment_Y}, $\hat{\mathbf{X}}_{d}$ can be written as
\begin{align}
	\hat{\mathbf{X}}_{d} =   \mathbf{X}_{d}+\mathbf{Z}_d', 
\end{align}
where $\mathbf{Z}_d'=\Re\left\{(\hat{\mathbf{H}}^H\hat{\mathbf{H}})^{-1}\hat{\mathbf{H}}^H\mathbf{Z}_d\right\}$. It can be shown that each column of $\mathbf{Z}_d'$ follows the distribution $\mathcal{N}\left(\mathbf{0},0.5\sigma_c^2(\hat{\mathbf{H}}^H\hat{\mathbf{H}})^{-1}\right)$. Moreover, from \eqref{eq_data}, each entry of $\mathbf{X}_d$ takes the form $\pm \sqrt{P_d}$. For the received symbol $r=s+n$, where $s\in\{\pm a\}$ and $n\sim\mathcal{N}(0,\tau)$, the log-likelihood ratio (LLR) is given by $\mathrm{LLR}(r)=\log\frac{p(r|s=+a)}{p(r|s=-a)}=\frac{2a}{\tau}\Re\{r\}$. Hence, the LLR for the $i$th user is calculated as
\begin{align}
\mathbf{f}_{d,i}=\dfrac{4\sqrt{P_d}}{\delta_{h_i}\sigma_c^2}\Re\{\hat{\mathbf{x}}_{d,i}\},\label{LLR_polar}
\end{align}
where $\delta_{h_i}$ is the $i$th diagonal element of the matrix $(\hat{\mathbf{H}}^H\hat{\mathbf{H}})^{-1}$, and $\hat{\mathbf{x}}_{d,i}$ is the $i$th row of $\hat{\mathbf{X}}_{d}$. Then, the LLR is fed to the polar list decoder. Motivated by \eqref{eq_appendedbits}, if the recovered bit sequence satisfies the CRC, the recovered polar submessage is appended to the detected pilot submessage (identified by the OMP algorithm) to obtain an estimate of the full ciphertext $\hat{\mathbf{c}}_i$. This estimate is then added to the set of successfully decoded messages, denoted by $\mathcal{S}$. 

\noindent \textbf{Step~3 (successive interference cancellation):}
For each bit sequence in the set $\mathcal{S}$, we regenerate the pilot and polar segments to construct the signal $\mathbf{x}^\prime_{p,i}$ of length $n_d + n_p$. We collect all $\mathbf{x}^\prime_{p,i}$ signals that are generated in the current and previous iterations in the row of the matrix $\mathbf{X}^\prime_p\in \mathbb{C}^{|\mathcal{S}|\times (n_d+n_p)}$, and estimate their corresponding channel coefficient vectors using LS as
\begin{align}
\hat{\mathbf{H}}=	\mathbf{Y}_{pp}{\mathbf{X}^\prime_p}^H\left(\mathbf{X}^\prime_p{\mathbf{X}^\prime_p}^H\right)^{-1},\label{Eq_chann_estLS}
\end{align} 
where $\mathbf{Y}_{pp} = \left[\mathbf{Y}_p,\mathbf{Y}_d\right]$. Note that the reason for re-estimation of the channel coefficient vectors is to obtain a more accurate estimation, because in
the channel estimation in Step~1, a length-$n_p$ pilot is used while in \eqref{Eq_chann_estLS}, the length-$(n_p+n_d)$ signal is served as pilot which gives a better estimation. Finally, the contribution of all successfully decoded messages in the current and previous iterations is removed from the received signal as
\begin{align}
	\mathbf{Y}_{pp}^\prime = \mathbf{Y}_{pp}-\hat{\mathbf{H}}\mathbf{X}^\prime_p.
\end{align}
The remaining received signal, \(\mathbf{Y}_{pp}^\prime\), is passed back to Step~1 for the next iteration. The iterative algorithm stops when no new messages are successfully decoded in an iteration. After the algorithm terminates, the set of successfully decoded ciphertexts, together with the corresponding estimated channel coefficients, are output by the iterative algorithm.

\subsubsection{Decoding Secret Key}
\label{Sec_Decodingsecretkey}
By substituting the estimated channel coefficient vectors obtained in \eqref{Eq_chann_estLS} into \eqref{Eq_est_yi}, an estimate of $\mathbf{y}_i$ is obtained. To generate the full LLR of length $n_s$ for the $i$th user for feeding to the LDPC decoder, the BS uses the estimated signal in \eqref{Eq_est_yi} to generate the LLR corresponding to the $S$ systematic symbols, shown by $	\mathbf{f}_{s,i}$, and the signal $\mathbf{Y}_k$ in \eqref{Eq_segment_Y} for generating the LLR corresponding to $n_s-S$ parity symbols, $\mathbf{f}_{p,i}$. The former is obtained using Appendix~\ref{Appndx_LLR} as
 \begin{align}
 	\mathbf{f}_{s,i} =[\nu_{i,1},\nu_{i,2},...,\nu_{i,S}],\label{LLR_system}
 \end{align}
 with 
 \begin{align}
	\nu_{i,j} &= \log\left( q_{u_{i,j}}\right)-\log\left( 1-q_{u_{i,j}}\right),
\end{align}
where $q_{u_{i,j}} = Q\!\left( \sqrt{\frac{2}{\sigma_u^2+\sigma_c^2\delta_{s_i}P_fM}}\,\hat{u}_{i,j} \right)$, $\delta_{s_i}$ is the $i$th diagonal entry of $\left(\mathbf{X}^\prime_p{\mathbf{X}^\prime_p}^H\right)^{-1}$, and $\hat{u}_{i,j}$ denotes the $j$th entry of $\hat{\mathbf{u}}_i$, which is an estimate of $\mathbf{u}_i$ in \eqref{eq_ui}, obtained as
\begin{align}
 	\hat{\mathbf{u}}_i = \left[\Re\{\hat{\mathbf{y}}_i \mathbf{C}_1\},\Im\{\hat{\mathbf{y}}_i \mathbf{C}_1\}\right] \in \mathbb{R}^{1 \times S}.\label{eq_uiPrime}
 \end{align}
In a similar way to obtain \eqref{LLR_polar}, we apply LS estimate on the $\mathbf{Y}_k$ in~\eqref{Eq_segment_Y}, for which the LLR corresponding to the parity symbols is calculated as
 \begin{align}
\mathbf{f}_{p,i}=\dfrac{4\sqrt{P_k}}{\delta_{h_i}\sigma_c^2}\Re\{\hat{\mathbf{x}}_{k,i}\},\label{LLR_ldpc}
 \end{align}
 where $\hat{\mathbf{x}}_{k,i}$ is the $i$th row of the following matrix
\begin{align}
\hat{\mathbf{X}}_k =  \Re\left\{ (\hat{\mathbf{H}}^H\hat{\mathbf{H}})^{-1}\hat{\mathbf{H}}^H\mathbf{Y}_k\right\}.\label{softParity}
\end{align} 
The full LLR is then obtained by appending LLRs in \eqref{LLR_system} and \eqref{LLR_ldpc} as
\begin{equation}
\mathbf{f}_{k,i} = \left[\mathbf{f}_{s,i},\mathbf{f}_{p,i}\right].\label{eq_llrMAIN}
\end{equation}
By feeding $\mathbf{f}_{k,i}$ to the LDPC decoder, an estimate of the secret key bits of the $i$th user, $\hat{\mathbf{s}}_i$, is recovered.

\begin{figure}
	\centering
	\includegraphics[width=\linewidth, trim=29 140 419 21, clip]{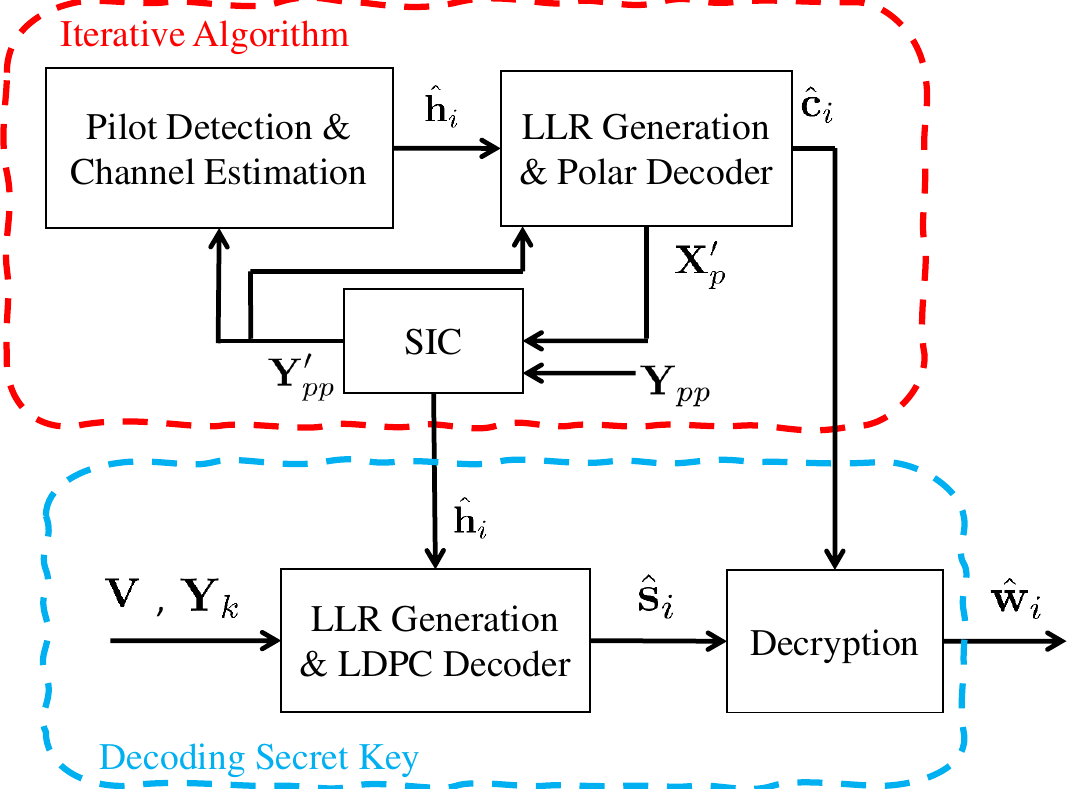}
\caption{{\small Block diagram of the receiver algorithm for the legitimate base station in the SURA system.}}
	\label{configuration}
\end{figure}

Finally, motivated by the encryption procedure in \eqref{eq_encrypted}, and using the estimated ciphertext obtained in Section~\ref{Sec_OMP} together with the estimated secret key from Section~\ref{Sec_Decodingsecretkey}, we obtain an estimate of the data bits of the \(i\)th user as

\begin{equation}
	\hat{\mathbf{w}}_i =   \hat{ \mathbf{c}}_i \oplus (\hat{\mathbf{s}}_i\mathbf{T} \bmod 2). \label{eq_decoded_databits}
\end{equation}
The receiving algorithm is detailed in Algorithm~\ref{algorithm2} and Fig.~\ref{configuration}.
\begin{algorithm}
	\caption{Receiver} 
	\label{algorithm2}
	\textbf{Iterative Decoding in Section~\ref{Sec_OMP}}:
	\begin{itemize}
	\item Step~1: Pilot detection \& channel estimation.
	\item Step~2: Polar decoding.
	\item Step~3: Channel re-estimation \& SIC.
	\end{itemize}
These three steps are iteratively repeated until no new successful decoding happens in an iteration. When the iterations stop, the algorithm outputs a set of estimated ciphertexts ${\hat{\mathbf{c}}_i}$.
		
	\textbf{Decoding Secret Key in Section~\ref{Sec_Decodingsecretkey}:}
	
	\begin{itemize}
		\item Feed LLR in~\eqref{eq_llrMAIN} to the LDPC decoder to obtain an estimate of the secret key, $\hat{\mathbf{s}}_i$.
		\item Decrypt the decoded messages $\hat{\mathbf{c}}_i$ using $\hat{\mathbf{s}}_i$ as in \eqref{eq_decoded_databits}.
	\end{itemize}
\end{algorithm}

\section{Theoretical Analysis}
\label{SecTheoretical}
In this section, we analyze the information-theoretic performance of the SURA scheme by deriving theoretical lower bounds on the PUPEs of the legitimate BS and the passive Eve. Specifically, for each user $i$, we first derive the information density at the legitimate receiver under assumptions favorable to its performance. Then, by applying a Gaussian approximation, we translate it into a lower bound on the block error probability for that user. The system's overall PUPE is subsequently obtained by averaging the per-user error probabilities across all active users. The same methodology is applied to the information density at a passive Eve to obtain a lower bound on the Eve's PUPE, which allows us to quantify the secrecy level of each individual transmission. We then analyze the additional overhead tolerated by the system when security measures are applied, relative to the original feedback-aided URA without security. These theoretical results provide benchmarks for the system's performance, secrecy level, and overhead.

For analytical tractability in deriving the theoretical PUPEs of the legitimate BS and the passive Eve, we make the following assumptions throughout this section. These assumptions are consistent with favoring the recovery performance of both receivers.  

\noindent\textbf{Assumptions}:
\begin{description}
    \item[(A1)] The information densities are assumed to be Gaussian distributed, characterized by their first and second moments.
    \item[(A2)] The iterative algorithm in Sec.~\ref{Sec_OMP} is assumed to perfectly recover all the ciphertexts for all users.
    \item[(A3)] For analyzing Eve's performance, we assume a single-user scenario with no interference, meaning that all other users' signals are perfectly removed at Eve's receiver. For the legitimate BS, no idealistic assumptions are made regarding interference.
    \item[(A4)] The parity symbols $\mathbf{x}_{k,i}$ are assumed to provide sufficient information to perfectly recover the secret key $\mathbf{s}_i$.
\item[(A5)] The elements of the parity vector $\mathbf{x}_{k,i}$ in \eqref{eq_keysegmentss} are independent and take values in $\{\pm \sqrt{P_k}\}$ with equal probability. The elements of the secret key $\mathbf{s}_i$ are i.i.d.\ Bernoulli$(0.5)$, taking values in $\{0,1\}$.
\end{description}
\subsection{Error Analysis for the Legitimate Receiver}
\label{section_TheoryLegReceiver}
We now focus on the legitimate receiver and derive the per-user information density, which is used to obtain a lower bound on the block error probability via a Gaussian approximation. The achievable information density for the $i$th user between the BS observations and the secret key can be expressed as
\begin{equation}
i_{\mathrm{BS,secret}} = i(\hat{\mathbf{x}}_{k,i}, \hat{\mathbf{u}}_i; \mathbf{s}_i ),
\end{equation}
where $i(\cdot)$ denotes the information density function. Here, $\hat{\mathbf{x}}_{k,i}$ and $\hat{\mathbf{u}}_i$ denote the estimates of the parity and systematic parts of the LDPC codeword for the $i$th user, as given by \eqref{softParity} and \eqref{eq_uiPrime}. These estimates are used to generate the corresponding LLRs for LDPC decoding. The above information density quantifies the amount of information about the secret key $\mathbf{s}_i$ that can be inferred from the decoder's soft observations in a single realization.

Using the chain rule for information density, it can be decomposed as
\begin{equation}
i_{\mathrm{BS,secret}}
= i(\hat{\mathbf{x}}_{k,i} ; \mathbf{s}_i \mid  \hat{\mathbf{u}}_i)
+ i(\hat{\mathbf{u}}_i; \mathbf{s}_i ),
\label{chainrule_density_i}
\end{equation}
where the first term captures the contribution of the parity symbol estimates $\hat{\mathbf{x}}_{k,i}$, while the second term represents the information about the secret key $\mathbf{s}_i$ provided by the estimate of the systematic part $\hat{\mathbf{u}}_i$.

Using assumptions (A1), and applying the Berry–Esseen theorem, the block error probability for the $i$th user can be written as
\begin{equation}
\label{eq:lowerbound_PUPE_nonuniform_explicit}
P_b^{(i)} =
Q\!\Bigg(
\frac{ I_x+I_u - S }
{\sqrt{ V_{2,x}+V_{2,u} }}
\Bigg)
+
\mathcal{O}\Bigg(
D
\Bigg),
\end{equation}
where the deviation term $D$ is defined as
\begin{align}
\label{eq:D_def}
D =
\frac{ V_{3,x}+V_{3,u} }
{ (V_{2,x}+V_{2,u})^{3/2} 
  \left[
    1 + \Big|
      \frac{I_x + I_u - S}{\sqrt{V_{2,x}+V_{2,u}}}
    \Big|^3
  \right]
},
\end{align}
with
\begin{subequations}
\label{momentsEq}
    \begin{align}
        I_x &= \mathbb{E}[\,i(\hat{\mathbf{x}}_{k,i} ; \mathbf{s}_i \mid  \hat{\mathbf{u}}_i)\,],\\
        I_u &= \mathbb{E}[\,i(\hat{\mathbf{u}}_i; \mathbf{s}_i )\,],\\
        V_{k,x} &= \mathbb{E}[\,|\,i(\hat{\mathbf{x}}_{k,i} ; \mathbf{s}_i \mid  \hat{\mathbf{u}}_i) - I_x\,|^k\,],\\
        V_{k,u} &= \mathbb{E}[\,|\,i(\hat{\mathbf{u}}_i; \mathbf{s}_i ) - I_u\,|^k\,].
    \end{align}
\end{subequations}
This formulation is justified by non-uniform Berry--Esseen bounds (e.g., \cite{ChenShao2001,Esseen1945}), which show that the approximation error decreases in the tails 
\(\Big|\frac{I_x + I_u - S}{\sqrt{V_{2,x}+V_{2,u}}}\Big|\) and is largest when the standardized deviation 
\(\frac{I_x + I_u - S}{\sqrt{V_{2,x}+V_{2,u}}}\) is close to zero, reflecting the improved accuracy of the Gaussian approximation in both low- and high-error regimes.

In the following, we derive the individual information density terms required to evaluate $P_b^{(i)}$ via these moments.

\vspace{3mm}

\noindent \textbf{Evaluation of $i(\hat{\mathbf{x}}_{k,i} ; \mathbf{s}_i \mid  \hat{\mathbf{u}}_i)$ (parity contribution to information density):}

\begin{mydef3}
\label{lemma1}
Consider the channel model
\begin{equation}
\mathbf{y} = \mathbf{g}v + \mathbf{n},
\end{equation}
where $v$ is a discrete random variable uniformly distributed over $\{-\alpha, +\alpha\}$. The vector $\mathbf{g}$ and noise $\mathbf{n}$ are defined according to the noise type:
\begin{itemize}
    \item {Real Gaussian noise (RGN):} $\mathbf{g} \in \mathbb{R}^M$, and $\mathbf{n} \sim \mathcal{N}(\mathbf{0}, \frac{\sigma_n^2}{2}\mathbf{I}_M)$.
    \item {Circularly symmetric complex Gaussian noise (CGN):} $\mathbf{g} \in \mathbb{C}^M$, and $\mathbf{n} \sim \mathcal{CN}(\mathbf{0}, \sigma_n^2 \mathbf{I}_M)$.
\end{itemize}
Then, the information density between $\mathbf{y}$ and $v$ conditioned on $\mathbf{g}$ can be expressed as
\begin{align}
i(v;\mathbf{y}\mid \mathbf{g})= \log_2 \frac{ 2 }{1+\exp\!\left( \frac{-4\alpha^2 \|\mathbf{g}\|^2 -4\alpha g}{\sigma_n^2} \right)}.
\end{align}
where $g =\Re\{\mathbf{g}^H\mathbf{n}\}\sim \mathcal{N}\!\left(0, \frac{\sigma_n^2}{2}\|\mathbf{g}\|^2\right)$,
\end{mydef3}
\begin{proof}
    See Appendix~\ref{appendixLemma1}.
\end{proof}
From \eqref{Eq_segment_Y} and \eqref{channelEstimation}, the received signal matrix corresponding to the key segment is expressed as
\begin{align}
    \mathbf{Y}_k&=\hat{\mathbf{H}}\mathbf{X}_k-\mathbf{Z}_{p}''\mathbf{X}_k+\mathbf{Z}_k.\label{eq_Yprime2}
\end{align} 
By substituting \eqref{eq_Yprime2} into \eqref{softParity}, and considering the real-valued structure of the signal matrix $\mathbf{X}_k$, the LS estimate of the parity part is obtained as
\begin{align}
\hat{\mathbf{X}}_k =  \mathbf{X}_k+\mathbf{Z}_i',\label{eq_34March5}
\end{align} 
where $\mathbf{Z}_i' = \Re\Big\{ (\mathbf{H}^H \mathbf{H})^{-1} \mathbf{H}^H (\mathbf{Z}_k-\mathbf{Z}_{p}''\mathbf{X}_k) \Big\}$. 
Looking at the structure of $\mathbf{Z}_{p}''$ in \eqref{eq_Z''} and since $\mathbb{E}\{\mathbf{X}_k\mathbf{X}_k^H\}=P_k(n_s-S)\mathbf{I}_{K_a}$, we can prove that each entry of $\mathbf{Z}_k-\mathbf{Z}_{p}''\mathbf{X}_k$ follows $\mathcal{CN}\left(0,\sigma_c^2\left(1+P_k\sum_{i=1}^{K_a}\delta_{s_i}\right)\right)$, where $\delta_{s_i}$ is the $i$th diagonal entry of $\left(\mathbf{X}^\prime_p{\mathbf{X}^\prime_p}^H\right)^{-1}$. Hence, each column of $\mathbf{Z}_i'$ is distributed as $\mathcal{N}\Big(\mathbf{0}, 0.5\, \sigma_c^2\left(1+P_k\sum_{i=1}^{K_a}\delta_{s_i}\right) \, \Re\{ (\mathbf{H}^H \mathbf{H})^{-1} \}\Big)$.

The entry in the $i$th row and the $j$th column of $\hat{\mathbf{X}}_k$ in \eqref{eq_34March5}, corresponding to user $i$ in the $j$th symbol, can be written as
\begin{equation}
\hat{x}_{k,i,j} = x_{k,i,j} + z_{i,j}',\label{eq36}
\end{equation}
where $x_{k,i,j} \in \left\{\pm \sqrt{P_k}\right\}$ denotes the $j$th entry of $\mathbf{x}_{k,i}$ in \eqref{eq_keysegmentss}, and 
\begin{align}
    z_{i,j}' \sim \mathcal{N}\left(0, 0.5\sigma_x^2 \right),\label{dist_z'}
\end{align}
where $\sigma_x^2=\sigma_c^2\left(1+P_k\sum_{i=1}^{K_a}\delta_{s_i}\right) \, \delta_{h_i}$, and $\delta_{h_i}$ denotes the $i$th diagonal element of $\Re\{ (\mathbf{H}^H \mathbf{H})^{-1} \}$.

Focusing on \eqref{eq36} and assumption~(A5), and noting that $x_{k,i,j}$ is conditionally independent of $\hat{\mathbf{u}}_i$ given $\hat{x}_{k,i,j}$, the model satisfies the conditions of Lemma~\ref{lemma1}. Hence, by applying this lemma, the conditional information density can be written as
\begin{align}
   i(\hat{x}_{k,i,j};x_{k,i,j} \mid \hat{\mathbf{u}}_i) = f_x(z_{i,j}'),\label{eq_perChuse_infodesnity}
\end{align}
where 
\begin{align}
    f_x(z)=\log_2 \frac{ 2 }{1+\exp\!\left( \frac{-4P_k  -4\sqrt{P_k} z}{\sigma_x^2} \right)}
\end{align}
From assumption~(A4), where the parity symbols $\mathbf{x}_{k,i}$ provide sufficient information to perfectly recover the secret key $\mathbf{s}_i$, we can write
\begin{align}
    i(\hat{\mathbf{x}}_{k,i} ; \mathbf{s}_i \mid  \hat{\mathbf{u}}_i) = i(\hat{\mathbf{x}}_{k,i}; \mathbf{x}_{k,i} \mid \hat{\mathbf{u}}_i).\label{Eq_infodensityEquality}
\end{align}
Furthermore, from assumption~(A5), the moments of $i(\hat{\mathbf{x}}_{k,i}; \mathbf{x}_{k,i} \mid \hat{\mathbf{u}}_i)$ can be expressed as $(n_s-S)$ times the moments of $i(\hat{x}_{k,i,j};x_{k,i,j} \mid \hat{\mathbf{u}}_i)$. Hence, using \eqref{momentsEq}, \eqref{eq_perChuse_infodesnity}, and \eqref{Eq_infodensityEquality}, and considering the distribution of $z_{i,j}'$ in \eqref{dist_z'}, we obtain
\begin{subequations}
\label{eq_momentsX}
\begin{align}
I_x &= (n_s-S)E_z,\\
 V_{k,x}&= (n_s-S)\int_{-\infty}^{\infty} |f_x(z) - E_z|^k 
\frac{e^{-\frac{z^2}{{\sigma_x^2}}}}{\sqrt{\pi {\sigma_x^2 }}}
\, \mathrm{d}z,
\end{align}
\end{subequations}
where 
$E_z =\mathbb{E}[f_x(z_{i,j}')]= \displaystyle\int_{-\infty}^{\infty}f_x(z)
\frac{e^{-\frac{z^2}{{\sigma_x^2 }}}}{\sqrt{\pi {\sigma_x^2}}}
\, \mathrm{d}z$.

\vspace{1mm}
\noindent \textbf{Evaluation of $ i(\hat{\mathbf{u}}_i; \mathbf{s}_i )$ (systematic contribution to information density):}

Let $\hat{u}_{i,j}$ and $s_{i,j}$ are the $j$th entries of $\hat{\mathbf{u}}_i$ and $\mathbf{s}_i$, respectively. The information density $i(\hat{u}_{i,j}; s_{i,j} )$ can be expressed as
\begin{subequations}
    \begin{align}
    i(\hat{u}_{i,j}; s_{i,j} ) &=  \log_2 \left( \dfrac{p_{{s}_{i,j} \mid \hat{u}_{i,j}}({s}_{i,j} \mid \hat{u}_{i,j})}{ p_{{s}_{i,j}}({s}_{i,j})} \right) \label{eq40c}\\
&= 1+ \log_2 \left( p_{{s}_{i,j} \mid \hat{u}_{i,j}}({s}_{i,j} \mid \hat{u}_{i,j}) \right), \label{eq40d}
\end{align}
\end{subequations}
where \eqref{eq40d} follows from assumption~(A5), which implies $p_{{s}_{i,j}} = 0.5$. From \eqref{conditionalProbabilities}, we can observe that $({s}_{i,j} | \hat{u}_{i,j})\sim$Bernoulli$(q_{u_{i,j}})$, where 
\begin{align}
    q_{u_{i,j}} = Q\!\left( \sqrt{\frac{2}{\sigma_u^2+\sigma_c^2\delta_{s_i}P_fM}}\,\hat{u}_{i,j} \right),\label{q_uij_eq}
\end{align}
and $\delta_{s_i}$ is the $i$th diagonal entry of $\left(\mathbf{X}^\prime_p{\mathbf{X}^\prime_p}^H\right)^{-1}$. Under assumption~(A5), and considering the element-wise nature of $F(\cdot)$, the moments of $i(\hat{\mathbf{u}}_i; \mathbf{s}_i )$ can be expressed as the sum of the moments of $i(\hat{u}_{i,j}; s_{i,j} )$ over all $j=1,2,...,S$. Therefore, substituting \eqref{eq40d} into \eqref{momentsEq}, and using the distribution of $({s}_{i,j} \mid \hat{u}_{i,j})$, we obtain 
\begin{subequations}
\begin{align}
I_u &= 1 - \sum_{j=1}^S H_b(q_{u_{i,j}}),\\[1mm]\nonumber
V_{k,u} 
&= \sum_{j=1}^S \Big[ q_{u_{i,j}} \left|\log_2 q_{u_{i,j}} + H_b(q_{u_{i,j}})\right|^k 
\\&\quad + (1-q_{u_{i,j}}) \left|\log_2 (1-q_{u_{i,j}}) + H_b(q_{u_{i,j}})\right|^k \Big],\label{V_uk}
\end{align}
\label{eq_momentsU}
\end{subequations}
where $H_b(p) = - p \log_2(p) - (1-p) \log_2(1-p)$ is the binary entropy function. Finally, the PUPE of the proposed SURA system is calculated as~\cite{Ahmadi2023Unsourced}
\begin{align}
	\zeta_b = \dfrac{1}{K_a}\sum_{i=1}^{K_a}P_b^{(i)},\label{pupeTheoryBS}
\end{align}
where $P_b^{(i)}$ is the error probability for each user obtained by plugging \eqref{eq_momentsX} and \eqref{eq_momentsU} into \eqref{eq:lowerbound_PUPE_nonuniform_explicit}. 

\subsection{Error Probability and Secrecy at the Eavesdropper}
From \eqref{Eq_uplink_eav}, the signals transmitted by users are also received by the passive Eve under different channel conditions. To assess the system’s security, we evaluate the information potentially leaked to Eve by computing the corresponding information density. Using a Gaussian approximation of the information density, we then calculate the theoretical block error rate for each user as decoded by Eve and average over all users to obtain the PUPE from Eve’s perspective. 

Under assumptions~(A2) and (A3) and using \eqref{Eq_uplink_eav} and \eqref{eq_transmitStrc}, the received signal at Eve corresponding to the key segment of the \(i\)th user can be written as
\begin{align}
        \mathbf{Y}_{E, k} = \mathbf{g}_i \mathbf{x}_{k,i}+\mathbf{N}_k,\label{eav_interfernceFree}
\end{align}
where $\mathbf{N}_k$ and $\mathbf{Y}_{E,k}$ denote the submatrices of $\mathbf{N}$ and $\mathbf{Y}_{\mathrm{Eve}}$ corresponding to the key segment. 

Let $x_{k,i,j}\in \{\pm\sqrt{P_k}\}$ be the $j$th element of $\mathbf{x}_{k,i}$ in~\eqref{eq_keysegmentss} and $\mathbf{y}_{e, k,j}$ is the $j$th column of $\mathbf{Y}_{E, k}$ in \eqref{eav_interfernceFree}, which can be written as
\begin{align}
        \mathbf{y}_{e, k,j} = \mathbf{g}_i x_{k,i,j}+\mathbf{n}_k,\label{eav_interfernceFree_vecs}
\end{align}
with $\mathbf{n}_k \in \mathcal{CN}(\mathbf{0}, \sigma_e^2 \mathbf{I}_M)$. Focusing on \eqref{eav_interfernceFree_vecs}, we observe that the model satisfies the assumptions of Lemma~\ref{lemma1}. Therefore, by applying this lemma, the information density corresponding to the leakage of $x_{k,i,j}$ to $\mathbf{y}_{e, k,j}$ can be expressed as
\begin{align}
i(x_{k,i,j}; \mathbf{y}_{e, k,j} \mid \mathbf{g}_i) = f_e(r), \label{leackedPerChUse}
\end{align}
where $r =\Re\{\mathbf{g}_i^H\mathbf{n}_k\}\sim \mathcal{N}\!\left(0, \frac{\sigma_e^2}{2}\|\mathbf{g}_i\|^2\right)$, and
\begin{align}
    f_e(r)=\log_2 \frac{ 2 }{1+\exp\!\left( \frac{-4P_k \|\mathbf{g}_i\|^2 -4\sqrt{P_k} r}{\sigma_e^2} \right)}.
\end{align}
Following the same procedure as for obtaining the moments in \eqref{eq_momentsX}, and using assumptions~(A4) and (A5), we can compute
\begin{subequations}
\label{eq_momentse}
\begin{align}
I_e &= \mathbb{E}[\,i(\mathbf{s}_i; \mathbf{Y}_{E, k}\mid\mathbf{g}_i)\,] = (n_s-S)E_e,\\\nonumber
 V_{k,e}&= \mathbb{E}[\,|\,i(\mathbf{s}_i; \mathbf{Y}_{E, k}\mid\mathbf{g}_i) - I_e\,|^k\,]\\
 &=(n_s-S)\int_{-\infty}^{\infty} |f_e(r) - E_e|^k 
\frac{e^{-\frac{r^2}{{\sigma_e^2\|\mathbf{g}_i\|^2}}}}{\sqrt{\pi {\sigma_e^2\|\mathbf{g}_i\|^2}}}
\, \mathrm{d}r,
\end{align}
\end{subequations}
where 
$E_e =\mathbb{E}[f_e(r)]= \displaystyle\int_{-\infty}^{\infty}f_e(r)
\frac{e^{-\frac{r^2}{{\sigma_e^2\|\mathbf{g}_i\|^2}}}}{\sqrt{\pi {\sigma_e^2\|\mathbf{g}_i\|^2}}}
\, \mathrm{d}r$.

Motivated by \eqref{eq:lowerbound_PUPE_nonuniform_explicit}, we employ a Gaussian approximation of the information density, together with the Berry--Esseen theorem, to compute the block error probability for the $i$th user as
\begin{equation}
P_e^{(i)} =
Q\!\Bigg(
\frac{ I_e - S }
{\sqrt{ V_{2,e} }}
\Bigg)
+\mathcal{O}\left(\frac{ V_{3,e} }
{ (V_{2,e})^{3/2} 
  \left[
    1 + \Big|
      \frac{I_e - S}{\sqrt{V_{2,e}}}
    \Big|^3
  \right]
}\right).
\end{equation}
Finally, to evaluate the overall performance of Eve, we average the block error probabilities over all users to obtain the equivalent of the PUPE for Eve, which is given by
\begin{align}
    \zeta_e = \frac{1}{K_a}\sum_{i=1}^{K_a}  \max \left\{
    P_e^{(i)},
    \; 1 - 2^{-S}
    \right\},\label{PUPE_EVE}
\end{align}
where the term $1 - 2^{-S}$ denotes an intrinsic error floor for each user, induced by the fact that all $2^S$ possible messages are equiprobable. This follows directly from assumption~(A5), since the elements of the secret key $\mathbf{s}_i$ are i.i.d.\ Bernoulli$(0.5)$.
\subsection{Extra Resource and Processing Overhead Due to Security}
\vspace{2mm}
\noindent In this subsection, we evaluate the additional cost introduced by the secure URA scheme relative to a conventional feedback-aided URA system. We quantify this overhead along four dimensions: (i) signaling, measured by the length of the additional key segment relative to the original transmission; (ii) energy consumption, given by the transmit power of the key segment; (iii) delay, due to waiting for feedback before transmission; and (iv) processing complexity at the BS for recovering the secret key. This analysis provides a clear assessment of the trade-offs between security and efficiency in URA systems.

From a system perspective, the overhead introduced by security can be quantified by comparing the proposed SURA with the same feedback-aided URA system without the key segment. In terms of signaling, the baseline system requires the feedback signal of length \(L\) from the base station and the transmissions of length \(n_p+n_d\) from each user, resulting in $L+K_a(n_p+n_d)$ channel uses. In the secure system, an additional key segment of length \(n_s-S\) is embedded in each user transmission, leading to a total signaling of $L + K_a(n_p + n_d + n_s - S)$. Hence, the normalized signaling overhead added by the security measures is written as
\begin{align}
   \tilde{S}= \dfrac{n_s - S}{ L/K_a+n_p + n_d}.\label{eq_additional_signling}
\end{align}
Correspondingly, the energy consumption follows a similar pattern, accounting for the transmit power in each phase. The normalized energy overhead added by the security measures is obtained as
\begin{align}
       \tilde{E}= \dfrac{P_k(n_s - S)}{P_fL/K_a+P_p n_p + P_d n_d }.\label{eq_additional_energy}
\end{align}
Regarding delay, the secure scheme introduces an additional waiting time equal to the feedback length $L$ before users can transmit, whereas the baseline feedback-aided URA does not incur this delay, as users transmit immediately once they are ready. Although this introduces a fixed delay into the system, the scheme remains suitable for large-scale connectivity scenarios, since the delay is independent of the number of active users.

We observe that the additional signaling overhead $\tilde{S}$, energy overhead $\tilde{E}$, and delay introduced by the security segment remain limited and do not depend on the number of active users $K_a$. This indicates that, even with the added security, the scheme preserves the low-overhead characteristics of URA, which are key advantages when considering URA as a replacement for conventional grant-based techniques that typically incur signaling overhead and delay proportional to the number of users connected to the network.

The computational complexity of the proposed scheme can be calculated as follows. To compute the complexity of the iterative algorithm, we evaluate its computational cost in the $k$th iteration with $K_r = K_a - k$ remaining codewords~\cite{Ahmadi2023Unsourced}. For the OMP algorithm, the computational complexity over $K_r$ iterations can be approximated as $\mathcal{O}\big(
K_r M n_p 2^{B_p} + K_r^4 + K_r^3 n_p + K_r^2 M n_p
\big)$; for the polar decoding of the $K_r$ users, the complexity is given by $\mathcal{O}\big(
K_r^2 M + K_r^3 + K_r M n_d + K_r^2 n_d + K_r n_d \log n_d
\big)$; and for the successive interference cancellation, the computational complexity in the $k$th iteration with $(k-1)$ successfully decoded users is approximated as $\mathcal{O}\big(k^3 + k^2 (n_p+n_d) + M k^2 + M k (n_p+n_d)\big)$. Finally, by summing the three per-iteration terms, assuming $n_p \approx n_d$, and applying the summation approximation, the total computational complexity of the iterative algorithm over $K_a$ iterations is obtained as
\begin{align}
\mathcal{C}_{\mathrm{iter}}=\mathcal{O}\big(
K_a^5 + n_p K_a^4 + M n_p K_a^3 + M n_p 2^{B_p} K_a^2
\big).\label{complexity_iter}
\end{align}
For decoding the secret key, the computational complexity of generating the LLRs corresponding to the parity part is dominated by the LS-based estimation of $\hat{\mathbf{X}}_k$ in~\eqref{softParity}, resulting in a complexity of $\mathcal{O}\big(K_a^3 + K_a^2 M + K_a M (n_s - S)\big)$. The computation of the LLRs for the systematic part has complexity $\mathcal{O}(S L)$, where $L > S/2$. Therefore, the overall computational complexity of decoding the secret key is 
 \begin{align}
\mathcal{C}_{\mathrm{key}}=\mathcal{O}\big(K_a^3+K_a^2M+K_aM (n_s-S)+L^2\big).\label{complexity_key}
\end{align}
From \eqref{complexity_iter} and \eqref{complexity_key}, we observe that the additional computational complexity due to the secure key, $\mathcal{C}_{\mathrm{key}}$, is dominated by the complexity of the original iterative algorithm, $\mathcal{C}_{\mathrm{iter}}$, and therefore has a negligible impact on the overall processing load at the receiver. This indicates that the extra computational effort introduced by the security measures is minimal, since the iterative algorithm for recovering the pilot and polar segments is executed regardless of whether security is applied, whereas the key segment is processed only when security is enabled.

\section{Numerical Results}
\label{Numerical}
In this section, we evaluate the performance of the proposed system through Monte Carlo simulations. The simulations use the following parameter values and system specifications. A Rayleigh fading channel model is assumed between each user and both the BS and Eve, i.e., each element of $\mathbf{g}_i$ and $\mathbf{h}_i$ is drawn from $\mathcal{CN}(0,1)$. The elements of the downlink signal matrix $\mathbf{V}$ are drawn from $\mathcal{CN}(0,1)$ and then scaled such that each column has squared norm $P_f M$. The noise variances at the BS, Eve, and the users are set to $\sigma_c^2 = \sigma_e^2 = \sigma_u^2 = 1$. The downlink signal length is $L = 20$. The numbers of bit sequences are $B = 100$, $B_r = 11$, and $S = 40$. The numbers of receive antennas at Eve and the BS are set to $E = M = 50$.

\begin{figure}[t!]
	 	\centering
        \includegraphics[width=.85\linewidth, trim=110 240 90 170, clip]{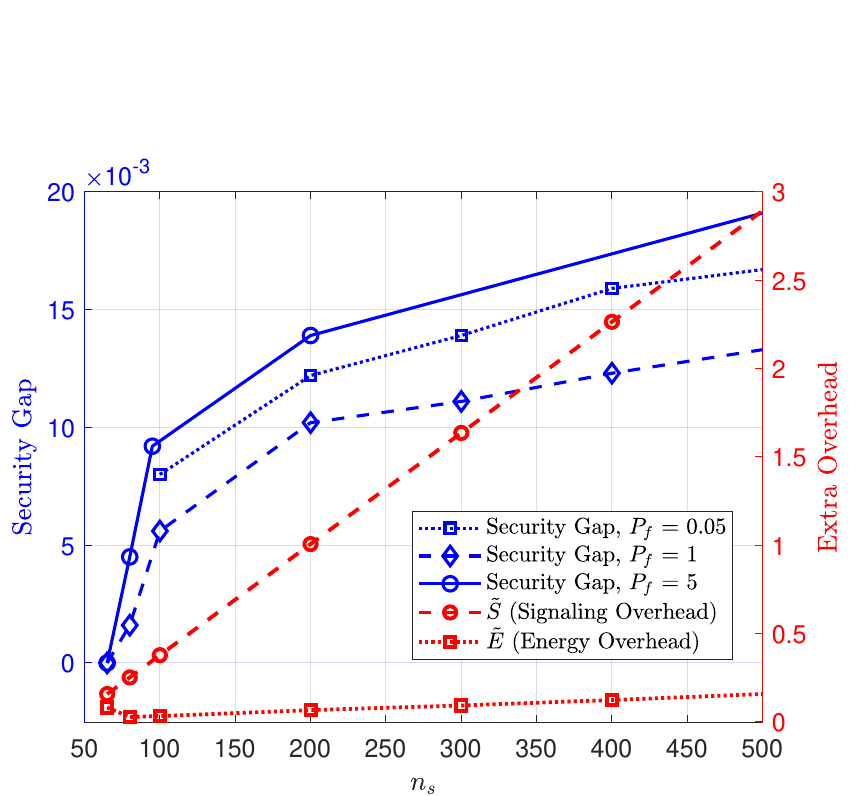}
\caption{{\small Impact of the LDPC-encoded secret key length $n_s$ on the security gap (left y-axis) and the normalized extra signaling and energy overheads $\tilde{S}$, $\tilde{E}$ (right y-axis).}}
\label{Fig_Overhead}
	 \end{figure}

In Fig.~\ref{Fig_Overhead}, we show the effect of increasing the length of the LDPC-encoded secret key, $n_s$, on the system's performance. Specifically, on the right y-axis, we plot the normalized extra signaling and energy overheads required by the security measures ($\tilde{S}$ and $\tilde{E}$ in \eqref{eq_additional_signling} and \eqref{eq_additional_energy}, respectively), while on the left y-axis, we plot the security gap (SG). \textcolor{black}{The SG is defined as the difference between the minimum required signal-to-noise ratio (SNR) at the legitimate receiver and the maximum tolerable SNR at the eavesdropper to achieve the target reliability levels, given by~\cite{Albashier2017,Matsumine2022}}

\begin{align}
\mathrm{SG}= \max\!\left(0,\dfrac{P_{\mathrm{tot}}^{\zeta_b}-P_{\mathrm{tot}}^{\zeta_e}}{(n_s - S+ n_p + n_d)\sigma_c^2}\right).\label{sasdfadfsa}
\end{align}
where $P_{\mathrm{tot}}^{\zeta_b}$ denotes the minimum signal energy required by the legitimate receiver in Sec.~\ref{Sec_poposdScheme} to achieve the target PUPE of $\zeta_b=0.05$, and $P_{\mathrm{tot}}^{\zeta_e}$ is the maximum signal energy such that the lower bound of Eve's PUPE in \eqref{PUPE_EVE} equals $\zeta_e=0.95$. These values are defined as
\begin{align}
P_{\mathrm{tot}}^{\zeta_b}=P_k^{\zeta_b}(n_s - S)+P_p^{\zeta_b} n_p + P_d^{\zeta_b} n_d,\\
P_{\mathrm{tot}}^{\zeta_e}=P_k^{\zeta_e}(n_s - S)+P_p^{\zeta_e} n_p + P_d^{\zeta_e} n_d.
\end{align}
Here, $P_l^{\zeta_b},\ l={k,p,d}$, are the minimum per-channel-use powers of the different segments required to achieve $\zeta_b$ at the legitimate BS, and $P_l^{\zeta_e},\ l={k,p,d}$, are the maximum per-channel-use powers required to satisfy $\zeta_e$ at Eve. For generating the results in Fig.~\ref{Fig_Overhead}, we set $K_a=20$, $B_p = 11$, $n_d = 128$, $n_p = 30$, $P_d^{\zeta_e}=P_d^{\zeta_b} =0.3$, and $P_p^{\zeta_e}=P_p^{\zeta_b} =0.5$, while $P_k^{\zeta_e}$ and $P_k^{\zeta_b}$ are optimized to satisfy the target PUPEs.

 From Fig.~\ref{Fig_Overhead}, it is evident that excessively increasing $n_s$ degrades the system’s security while also increasing the overhead, which is undesirable. On the other hand, for very small values of $n_s$, the system fails to achieve the target PUPE of $\zeta_b = 0.05$; for instance, this occurs when $n_s < 100$ for $P_f = 0.05$ and $n_s < 60$ for $P_f = 1$ and $ 5$. Hence, choosing $n_s$ too small deteriorates the performance at the legitimate base station. Therefore, a moderate and efficient value of $n_s$ should be selected based on the system parameters.

A similar trend can be observed for $P_f$. A moderate value is required, since either increasing or decreasing $P_f$ excessively leads to a larger security gap. This behavior follows from \eqref{eq_yhat}, which gives the BS estimate of the $i$th user's private signal. The downlink signal $\mathbf{V}$ appears in both the signal and noise terms, implying that increasing the downlink power $P_f$ scales both. The presence of $\mathbf{V}$ in the noise term is due to channel estimation errors; hence, improving the channel estimation accuracy reduces this effect and mitigates the impact of $P_f$ on the effective noise. Consequently, choosing $P_f$ either too large or too small degrades system performance by reducing the effective signal-to-noise ratio. Therefore, an optimized value of $P_f$ should be selected to achieve the best overall performance.

Another important observation from Fig.~\ref{Fig_Overhead} is that the additional signaling and energy overheads remain comparable to the baseline system without security measures (less than three times extra signaling and less than 0.2 extra energy for the plotted range of $n_s$), indicating that the SURA system maintains overheads within a reasonable range. This makes it a favorable choice over conventional grant-based multiple access in next-generation wireless networks with massive connectivity, as it still preserves the URA system's efficiency for massive numbers of users.

\begin{figure}[t!]
	 	\centering
        \includegraphics[width=.85\linewidth, trim=30 100 200 150, clip]{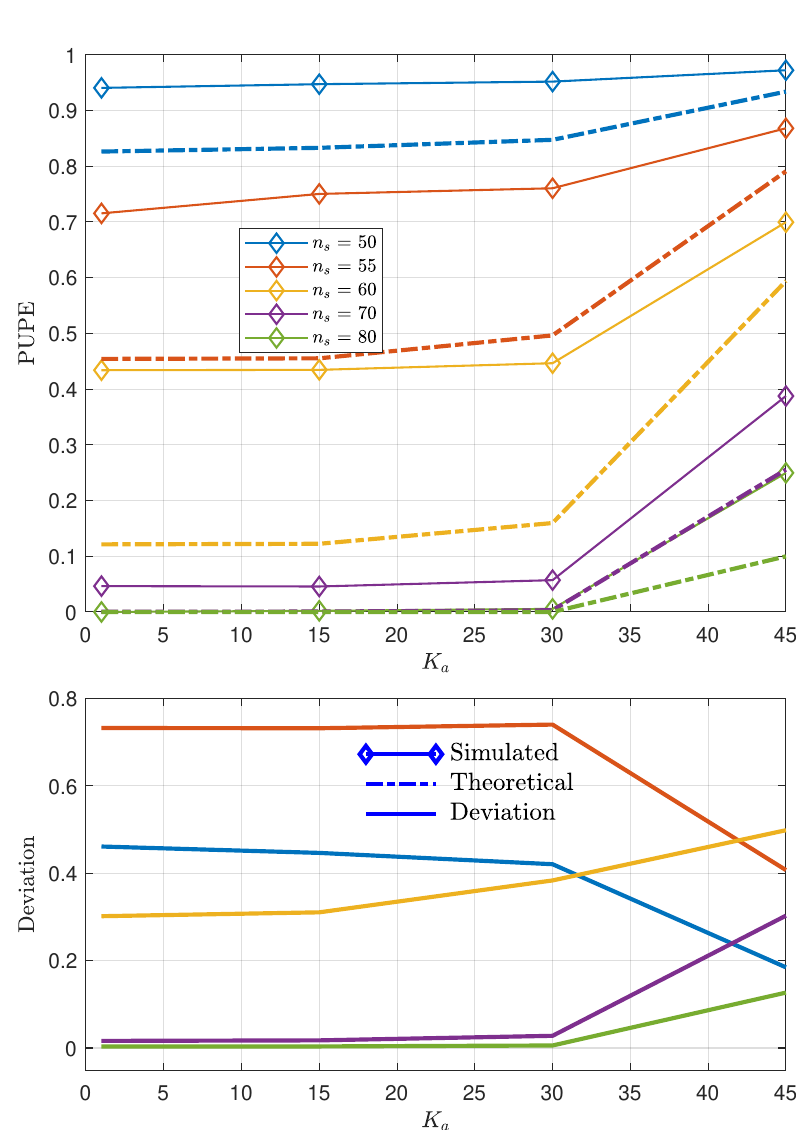}
\caption{{\small (Top) Comparison of theoretical and simulated PUPE versus the number of active users for different LDPC codeword lengths. (Bottom) Deviation term $D$ illustrating the non-uniform Berry--Esseen correction.}}
\label{Fig_theory}
	 \end{figure}

In Fig.~\ref{Fig_theory}, we examine the validity of the proposed theoretical derivations for the PUPE of the SURA system in \eqref{pupeTheoryBS} by comparing them with simulated PUPE results obtained from the SURA receiver described in Sec.~\ref{Sec_poposdScheme} under assumption~(A2). To generate the top subfigure, we neglect the deviation term in \eqref{eq:lowerbound_PUPE_nonuniform_explicit} for the theoretical PUPE and set $B_p=14$, $P_f = P_k = 0.15$, $n_d = 512$, and $n_p = 200$. In the bottom subfigure, we additionally plot the deviation term $D$ to provide further insight. The top subfigure shows that, despite the optimistic assumptions used in the theoretical derivation, the gap between simulated and theoretical curves is very small (less than 0.3). Furthermore, in regimes where the deviation is negligible (e.g., $n_s = 70, 80$ for $K_a \leq 30$), the theoretical and simulated curves perfectly match, further validating the theoretical model.

The top subfigure also illustrates that both theoretical and simulated PUPE values decrease as the LDPC codeword length $n_s$ increases, which is a consequence of the reduced channel coding rate. Another observation is that although PUPE increases with the number of active users $K_a$, this increase is smaller for lower values of $n_s$. This behavior occurs because secret key recovery in the systematic part is independent of $K_a$, while only the parity part is affected by $K_a$. Therefore, for small $n_s$, where the parity segment length $n_s-S$ is short, the recovery performance is less influenced by the parity part and hence less dependent on $K_a$.

%
\section{Conclusion}
\label{concluson}
This work has presented a secure communication framework for unsourced random access by incorporating physical layer security into a feedback-aided URA system without modifying its structure or operational characteristics. In the proposed design, each user extracts a private observation from the feedback signal broadcast by the base station, generates a secret key, and uses it to encrypt its data. A complete transmitter and receiver architecture was developed to support these operations. Furthermore, a comprehensive theoretical analysis was conducted to evaluate the performance of both the legitimate base station and a passive eavesdropper, as well as to quantify the additional overhead introduced by the security measures.
\begin{appendices}
\section{LLR Calculation for Systematic Symbols}
\label{Appndx_LLR}
Substituting \eqref{Eq_segment_Y} into \eqref{Eq_chann_estLS}, we have 
\begin{align}
\hat{\mathbf{H}}=\mathbf{H}+	\mathbf{Z}_{p}'',\label{channelEstimation}
\end{align} 
where 
\begin{align}
    \mathbf{Z}_{p}''=\mathbf{Z}_{p}'{\mathbf{X}^\prime_p}^H\left(\mathbf{X}^\prime_p{\mathbf{X}^\prime_p}^H\right)^{-1}\sim\mathcal{CN}\left(\mathbf{0},\sigma_c^2\left(\mathbf{X}^\prime_p{\mathbf{X}^\prime_p}^H\right)^{-1}\right)\label{eq_Z''}.
\end{align}
with $\mathbf{Z}_{p}'=[\mathbf{Z}_{p},\mathbf{Z}_{d}]$. To obtain this equation, we assume that all the active codewords are correctly decoded, i.e., $\mathbf{Y}_{pp}=\mathbf{H}\mathbf{X}^\prime_p+\mathbf{Z}_{p}'$.
Therefore, the estimated channel for the $i$th user can be written as
\begin{align}
    \hat{\mathbf{h}}_i=\mathbf{h}_i+\mathbf{z}_{p,i}'',\label{EstChannel}
\end{align}
where $\mathbf{z}_{p,i}''$ is the $i$th column of $\mathbf{Z}_{p}''$, hence $\mathbf{z}_{p,i}''\sim\mathcal{CN}\left(\mathbf{0},\sigma_c^2\delta_{s_i}\mathbf{I}_M\right)$ with $\delta_{s_i}$ beging the $i$th diagonal entry of $\left(\mathbf{X}^\prime_p{\mathbf{X}^\prime_p}^H\right)^{-1}$.
\vspace{1mm}

From \eqref{Eq_downlink} and \eqref{Eq_est_yi}, and \eqref{EstChannel}, we have
\begin{align}
	\nonumber  \hat{\mathbf{y}}_i &= \mathbf{h}_i^T\mathbf{V}+{\mathbf{z}_{p,i}''}^T\mathbf{V}\\    
	&= \mathbf{y}_i+\mathbf{o}_i', \label{eq_yhat}
\end{align}
where $\mathbf{o}_i'={\mathbf{z}_{p,i}''}^T\mathbf{V}-\mathbf{o}_i\sim \mathcal{CN}(0,\sigma_o^2\mathbf{I}_L)$ with $\sigma_o^2=\sigma_u^2+\sigma_c^2\delta_{s_i}P_fM$. An estimation of $\mathbf{u}_i$ in \eqref{eq_ui} can be obtained as
\begin{align}
     	\hat{\mathbf{u}}_i = \left[\Re\{\hat{\mathbf{y}}_i \mathbf{C}_1\},\Im\{\hat{\mathbf{y}}_i \mathbf{C}_1\}\right] \in \mathbb{R}^{1 \times S}.\label{eq_uiHat}
\end{align}
Plugging \eqref{eq_yhat} into \eqref{eq_uiHat}, we obtain
\begin{align}
    \mathbf{u}_i = \hat{\mathbf{u}}_i + \mathbf{e}_i, \label{u_hat_detailed}
\end{align}
where $\mathbf{e}_i = \big[\Re\{\mathbf{o}_{i}' \mathbf{C}_1\}, \; \Im\{\mathbf{o}_{i}' \mathbf{C}_1\}\big]$.  
Since $\mathbf{C}_1^H \mathbf{C}_1 = \mathbf{I}_{0.5S}$, it follows that $\mathbf{e}_i \sim \mathcal{N}\big(\mathbf{0}, 0.5\sigma_o^2 \mathbf{I}_S\big)$. From \eqref{u_hat_detailed} and \eqref{eq_secretKey}, we obtain
\begin{align}
    s_{i,j} = F(\hat{u}_{i,j}+e_{i,j}) =
    \begin{cases}
        1, & \hat{u}_{i,j} + e_{i,j} > 0,\\
        0, & \hat{u}_{i,j} + e_{i,j} \le 0,
    \end{cases}
    \label{Eq_sij}
\end{align}
where $s_{i,j}$, $\hat{u}_{i,j}$, and $e_{i,j}$ are the $j$th entries of $\mathbf{s}_{i,j}$, $\hat{\mathbf{u}}_{i,j}$, and $\mathbf{e}_{i,j}$, respectively. Hence, the conditional probability of $s_{i,j}$ given $\hat{u}_{i,j}$ is
\begin{equation}
\label{conditionalProbabilities}
p_{{s}_{i,j} \mid \hat{u}_{i,j}}({s}_{i,j} \mid \hat{u}_{i,j})  =
\begin{cases}
1 - q_{u_{i,j}}, & s_{i,j} = 1,\\[1ex]
q_{u_{i,j}}, & s_{i,j} = 0,
\end{cases},
\end{equation}
where $q_{u_{i,j}}=Q\!\left( \sqrt{\frac{2}{\sigma_o^2}}\hat{u}_{i,j} \right)$. Then, the LLR corresponding to the $j$th symbol of the systematic part is computed as
\begin{align}
	\nu_{i,j} &= \log\left(\dfrac{p_{{s}_{i,j} \mid \hat{u}_{i,j}}({s}_{i,j} =0\mid \hat{u}_{i,j})}{p_{{s}_{i,j} \mid \hat{u}_{i,j}}({s}_{i,j} =1\mid \hat{u}_{i,j})}\right).\label{eq4-}
\end{align}
 Using \eqref{conditionalProbabilities}, the LLR in \eqref{eq4-} can be written as
\begin{align}
	\nu_{i,j} &= \log\left( q_{u_{i,j}}\right)-\log\left( 1-q_{u_{i,j}}\right).
\end{align}
\section{Proof of Lemma~\ref{lemma1}}
\label{appendixLemma1}
For the model in Lemma~\ref{lemma1}, the distribution of $\mathbf{y}$ conditioned on $v$ and $\mathbf{g}$ is
\begin{align}
p_{\mathbf{y}\mid v,\mathbf{g}}(\mathbf{y}\mid v,\mathbf{g})
=
\begin{cases}
\frac{1}{(\pi \sigma_n^2)^{M/2}}
\exp\!\left(-\frac{\|\mathbf{y} - \mathbf{g}v\|^2}{\sigma_n^2}\right),
& \text{RGN}, \\[1em]
\frac{1}{(\pi \sigma_n^2)^M}
\exp\!\left(-\frac{\|\mathbf{y} - \mathbf{g}v\|^2}{\sigma_n^2}\right),
& \text{CGN}.
\end{cases}
\end{align}
Using this distribution and $p_v(v)=0.5$, the information density between $v$ and $\mathbf{y}$ conditioned on $\mathbf{g}$ can be rewritten as
\begin{subequations}
    \begin{align}
i(v; \mathbf{y}\mid \mathbf{g}) 
&= \log_2 \frac{p_{\mathbf{y}\mid v,\mathbf{g}}(\mathbf{y}\mid v,\mathbf{g})}
{p_{\mathbf{y}\mid \mathbf{g}}(\mathbf{y}\mid \mathbf{g})}\\
&= \log_2 \frac{p_{\mathbf{y} \mid v,\mathbf{g}}(\mathbf{y} \mid v,\mathbf{g})}{\frac{1}{2} \sum_{v' \in \{\pm \alpha \}}  p_{\mathbf{y} \mid v',\mathbf{g}}(\mathbf{y} \mid v',\mathbf{g})}\\
&= \log_2 \frac{ 2 \exp\!\left( - \frac{\|\mathbf{y} - v\mathbf{g}\|^2}{\sigma_n^2} \right)}{\sum_{v' \in \{\pm \alpha \}} \exp\!\left( - \frac{\|\mathbf{y} - v'\mathbf{g}\|^2}{\sigma_n^2} \right)} \\
&= \log_2 \frac{ 2 \exp\!\left( - \frac{\|\mathbf{n}\|^2}{\sigma_n^2} \right)}{\sum_{v' \in \{\pm \alpha \}} \exp\!\left( - \frac{\|\mathbf{n} + (v - v')\mathbf{g}\|^2}{\sigma_n^2} \right)} \\
&= \log_2 \frac{ 2 }{\sum_{v' \in \{\pm \alpha \}}\exp\!\left( - \frac{(v - v')^2 \|\mathbf{g}\|^2 + 2(v - v') g}{\sigma_n^2} \right)}\\
&=  \left\{\begin{matrix}
 f(g)& if v= \alpha  \\
 f(-g)&  if v=- \alpha \\
\end{matrix}\right.\\
&=  f(g),\label{info_denfianl}
\end{align}
\end{subequations}
where $g =\Re\{\mathbf{g}^H\mathbf{n}\}\sim \mathcal{N}\!\left(0, \frac{\sigma_n^2}{2}\|\mathbf{g}\|^2\right)$, and
\begin{align}
    f(g) = \log_2 \frac{ 2 }{1+\exp\!\left( \frac{-4\alpha^2 \|\mathbf{g}\|^2 -4\alpha g}{\sigma_n^2} \right)}.
\end{align}
In \eqref{info_denfianl}, we use the fact that $f(g)$ and $f(-g)$ have the same distribution, since $g$ is symmetrically distributed around zero.
\end{appendices}

\end{document}